\documentclass[prd,showpacs,superscriptaddress,twocolumn,
floatfix,preprintnumbers,altaffilletter]{revtex4}

\usepackage{longtable}
\usepackage{graphicx}
\usepackage{amsmath,amssymb}
\usepackage{color}
\usepackage{units}
\usepackage{epstopdf}

\newcommand{\beq}{\begin{equation}}
\newcommand{\eeq}{\end{equation}}
\newcommand{\bdm}{\begin{displaymath}}
\newcommand{\edm}{\end{displaymath}}

\definecolor{Gray}{gray}{0.9}

\graphicspath{{./plots/}}
\begin{document}


\title{Detecting compact binary coalescences with seedless clustering}
\author{M Coughlin}
\email{coughlin@physics.harvard.edu}
\affiliation{Department of Physics, Harvard University, Cambridge, MA 02138, USA}
\author{E Thrane}
\affiliation{LIGO Laboratory, California Institute of Technology, MS 100-36,
Pasadena, CA, 91125, USA}
\author{N Christensen}
\affiliation{Physics and Astronomy, Carleton College, Northfield, Minnesota 55057}
\date{\today}

\begin{abstract}

Compact binary coalescences are a promising source of gravitational waves for second-generation interferometric gravitational-wave detectors. 
Although matched filtering is the optimal search method for well-modeled systems, alternative detection strategies can be used to guard against theoretical errors (e.g., involving new physics and/or assumptions about spin/eccentricity) while providing a measure of redundancy.
In previous work, we showed how ``seedless clustering'' can be used to detect long-lived gravitational-wave transients in both targeted and all-sky searches.
In this paper, we apply seedless clustering to the problem of low-mass ($M_\text{total}\leq10M_\odot$) compact binary coalescences for both spinning and eccentric systems.
We show that seedless clustering provides a robust and computationally efficient method for detecting low-mass compact binaries.

\end{abstract}

\maketitle

\section{Introduction}\label{sec:Intro}
Compact binary coalescences (CBCs) of black holes (BHs) and/or neutron stars (NSs) are a likely source of gravitational waves (GWs) \cite{S6Highmass,S6Lowmass,0264-9381-27-17-173001}.
CBC events include binary neutron stars (BNSs), neutron-star black holes (NSBHs), and binary black holes (BBHs).
As a CBC passes through its inspiral and merger stage, it generates GWs which sweep upward in frequency and strain amplitude through the sensitive band of GW detectors.
The detection of GWs from CBCs will provide information about the populations of compact objects in the universe, elucidate the properties of strong field gravity, and provide a means to test general relativity.

Here, we focus on relatively low-mass binaries ($M_\text{total}\leq10M_\odot$).
There are two reasons for restricting our attention to this region of parameter space.
First, the rate of low-mass CBCs is less subject to theoretical uncertainty than high stellar-mass binary BHs and intermediate-mass BBH.
Second, we are interested in long-lived signals ($\approx54$--$\unit[270]{s}$), which appear as curved tracks in spectrograms of GW strain power, and therefore provide an appealing target for seedless clustering~\cite{PhysRevD.88.083010,stochsky} (described in greater detail below).

Searches for CBCs often use matched filtering, which requires precise knowledge of astrophysical waveforms.
(Excess power searches are also used, especially for high-mass systems associated with shorter signals; see, e.g.,~\cite{imbh}.)
Since CBCs are, for the most part, well-modeled systems, matched filtering provides an essentially optimal strategy for detecting compact binaries.
However, there are several reasons why it is useful to consider alternative detection strategies.

{\em Independent verification.}
Alternative methods can provide independent verification of detections by matched filter pipelines, thereby increasing confidence in the veracity of a result (of course, because seedless clustering will be less sensitive than matched filtering searches in most cases, non-detection by seedless clustering is not a concern either). Although there is some redundancy provided by the multiple implementations of matched filtering used in current searches, seedless clustering provides a very different approach to gravitational-wave detection and detection by both methods potentially indicates the robustness of the result.

{\em Visualization of the GW signal.}
In general, advanced detector CBC events are expected to be buried in noise to the extent that it will be difficult to see by eye their signature in a time series or strain auto-power spectrogram.
Here we show that, by coherently combining the output of multiple detectors, CBCs can be visualized as faint but visible arcs on ``radiometric spectrograms''---especially when the eye is guided by the reconstructed track of a search algorithm.
(The curious reader is encouraged to skip ahead to Fig.~\ref{fig:EBNS} for an example of a radiometric spectrogram.)

Visualizing the signal helps confirm that the detected signal looks like one expects.
The coherent combination also allows for confirmation that the parameter estimation of the signals, including the direction, masses, and time of coalescence, are all consistent ``by eye'' with the radiometric spectrogram.
For example, an error in the reconstructed CBC direction creates characteristic stripes; see~\cite{stochsky}.
If, on the other hand, the masses are incorrect, the reconstructed track will have the wrong frequency evolution as a function of time.

{\em Data processing corner cases.}
Real-world GW searches require design choices, which take into account the complicated nature of GW detectors.
Detector performance is non-stationary, the noise contains non-Gaussian ``glitches,'' and data-taking is sometimes interrupted by lock-loss, just to name a few relevant effects.
As a result, workarounds are employed, e.g., to estimate background, to discard noisy data, and to handle gaps. Matched filtering \cite{PhysRevD.87.024033} and seedless clustering \cite{stamp_glitch} have different ways of performing these tasks.
In general, these technical details are (by design) not important factors in determining the average sensitivity of a search. 
However, by employing multiple search methods we can guard against individual events falling between the cracks.

\begin{figure*}[t]
 \includegraphics[width=3.5in]{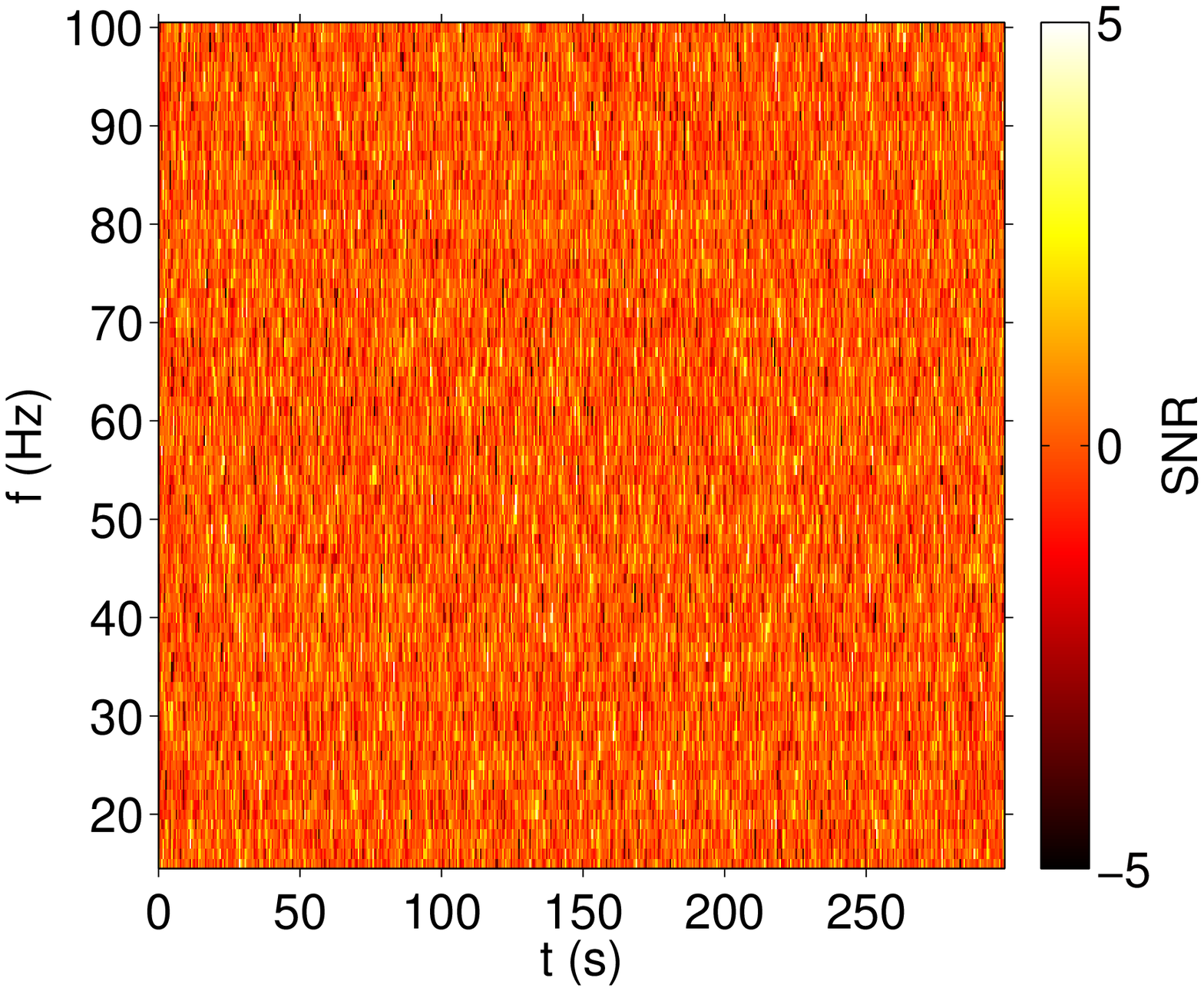}
 \includegraphics[width=3.5in]{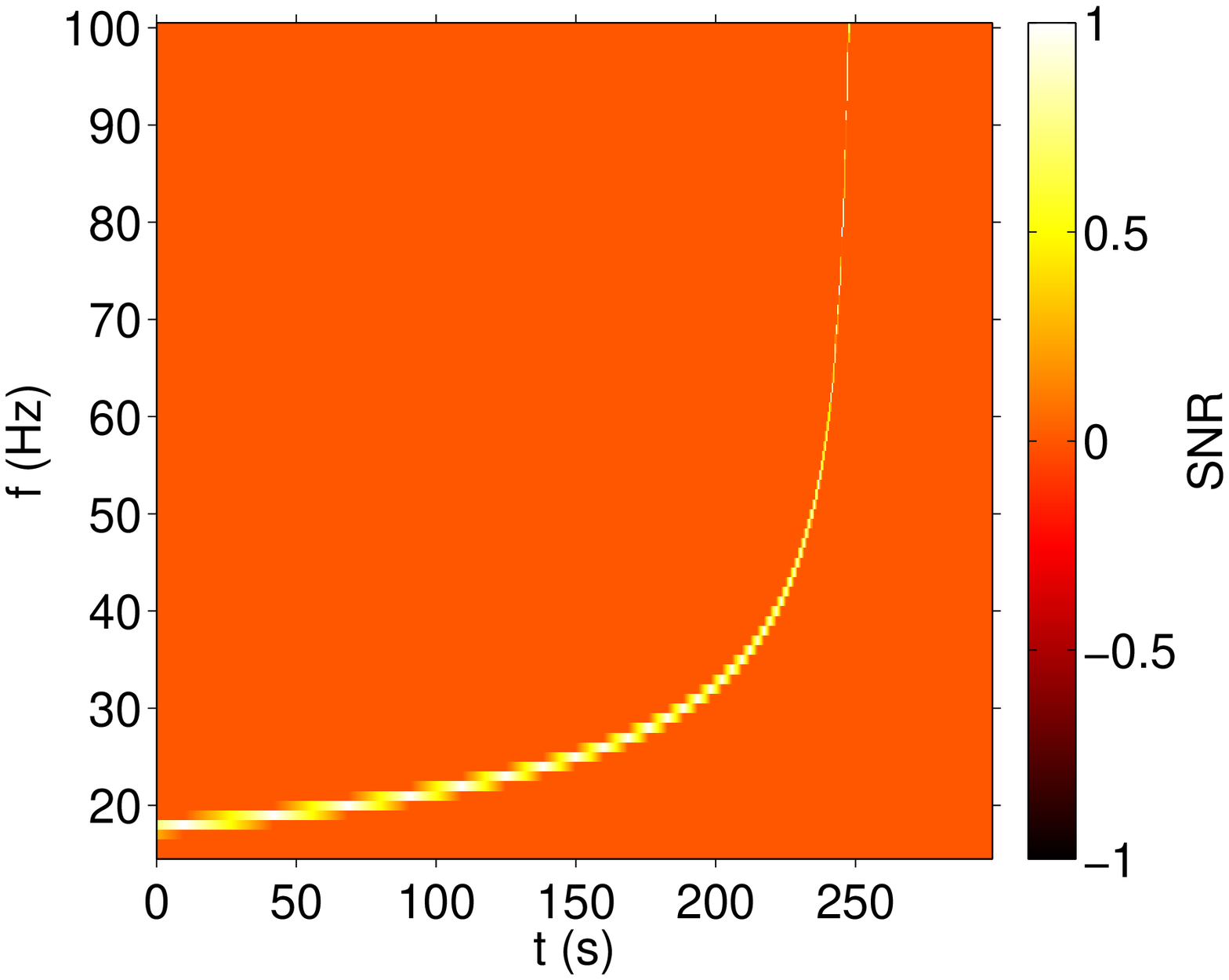}\\
 \caption{
   The plot on the left shows $\rho(t;f)$ during a recent LIGO engineering run, in which data from a LIGO subsystem---not sensitive to GW strain---is recolored to produce semi-realistic detector noise.
   On the right is the seedless clustering recovery, which is able to to detect the injected signal with high confidence $\text{FAP}<0.1\%$, despite the relatively poor data quality.
   (Though it is not immediately apparent from these plots, five segments are identified as characteristic of non-stationary noise~\cite{stamp_glitch}.)
 The signal is recovered with $\text{FAP}<0.1\%$.
 }
 \label{fig:ER5}
\end{figure*}

An example of a possible data processing corner case is shown in Fig.~\ref{fig:ER5}. This event was identified correctly by both matched filtering and seedless clustering.
The left-hand panel shows $\rho(t;f)$ obtained using ``engineering-run'' data\footnote{The data in question are from LIGO Engineering Run 5, collected during 2014.}, in which data from a LIGO sub-system, in this case the pre-stabilized laser, is recolored to match the Advanced LIGO noise curve.
Such engineering run data does not contain astrophysically useful strain measurements, but is nonetheless useful for its non-Gaussian noise characteristics.
The data contains five segments consistent with non-Gaussian noise and would be removed in a search.
Nonetheless, despite these noise artifacts, it is still possible to detect a simulated binary neutron star signal.
The right-hand panel shows the reconstructed signal, obtained with the seedless clustering algorithm we describe below.

{\em Waveform uncertainties.}
Theoretical errors in matched filter waveforms can arise from the computational limitations and/or imperfect approximations.
Due to computational limitations, most CBC searches so far use template banks composed of non-spinning, non-eccentric waveforms, which are less computationally challenging than search with spin and eccentricity.
High-spin systems take longer to simulate with numerical relativity and the addition of extra spin parameters creates larger, more unwieldy template banks. Searches that ignore spin can suffer significant losses in sensitivity~\cite{NSBHSpin}; the cases considered here have between a 23-36\% match, which is maximized over time and phase, between the spinning and non-spinning waveforms; these numbers increase to at least 97\% when maximized over mass as well.
When spin is included, it is often assumed that the spins are aligned in order to make the calculation more tractable.
Even so, the inclusion spin effects can lead to a factor of two increase in sensitive volume~\cite{SteveP:Thesis:2014}; see also~\cite{PhysRevD.86.084017,PhysRevD.89.024003}.
Main sequence binaries circularize by the time they enter the sensitive frequency band of terrestrial detectors~\cite{PhysRevD.81.024007}.
However, dynamical capture may produce gravitational waves from highly eccentric binaries~\cite{ScatteringBlackHoles,PhysRevD.81.024007,PhysRevD.78.064069,escidoc:795602,Gold:2011df}. The cases considered here have a less than 1\% match between the eccentric and circular waveforms of equivalent mass. These numbers increase to between 20-60\% when maximized over mass. This highlights the difficulty of detecting them using a template bank composed of circular templates.
Imperfect assumptions about eccentricity and spin may therefore create openings for ostensibly sub-optimal detection strategies.

{\em New physics.}
One can also imagine significant waveform errors due to the existence of new or unforeseen physics.
For example, Piro raised concerns about the effects of magnetic interactions in BNS~\cite{piro_mag}.
While these magnetic interactions were subsequently shown to be ignorable~\cite{dong} for GW astronomy, one can imagine a comparable source of theoretical error.
More speculatively, non-standard theories of gravitation can lead to modifications of the waveform \cite{PhysRevD.89.084005,PhysRevD.87.081506}.

Thus, there are many reasons why it is worth considering alternatives to matched filtering.
One common alternative technique for detection of GW transients is to search for excess power in spectrograms (also called frequency-time $ft$-maps) of GW detector data \cite{X-Pipeline,CoherentWaveBurst,STAMP}.
This method casts GW searches as pattern recognition problems.

Previous work has shown how ``seedless clustering'' can be used to perform sensitive searches for long-lived transients \cite{PhysRevD.88.083010,stochsky}.
The idea of seedless clustering is to integrate the signal power along spectrogram tracks chosen to capture the salient features of a wide class of signal models.
Seedless clustering calculations are embarrassingly parallel, and so the technique benefits from the recent proliferation of highly parallel computing processors including graphical processor units and multi-core central processing units.
Previous papers~\cite{PhysRevD.88.083010,stochsky} have pointed out that seedless clustering algorithms might be useful for CBC detection/confirmation.

In this work, we apply the seedless clustering formalism to efficiently search for CBC signals.
In section \ref{sec:SeedlessClustering}, we review the basics of seedless clustering.
We show how the formalism of~\cite{PhysRevD.88.083010,stochsky} can be tuned to more sensitively detect CBC signals.
In section \ref{sec:Results}, we determine the sensitivity of seedless clustering algorithms (with different levels of tuning) to CBC waveforms.
We conclude with a discussion of topics for further study in section \ref{sec:Conclusion}.

\section{Seedless clustering for chirps}
\label{sec:SeedlessClustering}
Searches for unmodeled GW transients typically begin with spectrograms proportional to GW strain power.
The pixels of these spectrograms are computed by dividing detector strain time series in segments and computing Fourier transform of the segments.
The Fourier transform of the strain data from detector $I$ for the segment with a mid-time of $t$ is denoted $\tilde{s}_I(t;f)$.
For the results presented here, we use $50\%$-overlapping, Hann-windowed segments with duration of $\unit[1]{s}$.
The frequency resolution is $\unit[1]{Hz}$.

Searches for long-duration GW transients in particular use the cross-correlation of two GW strain channels from spatially separated detectors to construct $ft$-maps of cross-power signal-to-noise ratio \cite{STAMP}:
\begin{equation}
  \rho(t;f|\hat\Omega) = \text{Re}\left[
    \lambda(t;f)
    e^{2\pi i f \Delta\vec{x}\cdot\hat\Omega/c}
    \tilde{s}_I^*(t;f) \tilde{s}_J(t;f)
    \right] .
\end{equation}
Here, $\hat\Omega$ is the direction of the GW source, $\Delta\vec{x}$ is a vector describing the relative displacement of the two detectors, $c$ is the speed of light, and $e^{2\pi i f \Delta\vec{x}\cdot\hat\Omega/c}$ is a direction-dependent phase factor, which takes into account the time delay between the two detectors.
The $\lambda(t;f)$ term is a normalization factor, which employs data from neighboring segments to estimate the background at time $t$:
\begin{equation}
    \lambda(t;f) = \frac{1}{\cal N} \sqrt{\frac{2}{P'_I(t;f) P'_J(t;f)}} .
\end{equation}
$P'_I(t;f)$ and $P'_J(t;f)$ are the auto-power spectral densities for detectors $I$ and $J$ in the segments neighboring $t$.
For additional details, see~\cite{STAMP,PhysRevD.88.083010,stochsky}

GWs appear as tracks or blobs in the $ft$-maps.
The morphology of the the GWs are dependent on the signal.
CBC signals appear as chirps of increasing frequency.
Clustering algorithms are used to identify clusters of pixels $\Gamma$ likely to be associated with a GW signal.
The total signal-to-noise ratio for a cluster of pixels can be written as a sum of over $\rho(t;f|\hat\Omega)$:
\begin{equation}\label{eq:sum}
  \text{SNR}_\text{tot} \equiv
  \frac{1}{N^{1/2}}
  \sum_{\left\{t;f\right\}\in\Gamma} \rho(t;f|\hat\Omega) ,
\end{equation}
where $N$ is the number of pixels in $\Gamma$.

Different clustering algorithms employ different methods for choosing $\Gamma$.
Seed-based algorithms connect statistically significant seed pixels to form clusters~\cite{PhysRevD.88.083010,burstegard}.
In seedless clustering algorithms~\cite{PhysRevD.88.083010}, $\Gamma$ is chosen from a bank of parametrized frequency-time tracks.
Each such track is referred to as a ``template.''
Calculations for many templates can be carried in parallel, which facilitates rapid calculations on multi-core devices such as graphical processor units (GPUs).

It must be noted that our templates are different from matched-filtering templates.
Seedless clustering templates describe the morphology of a power spectrogram track whereas matched filter templates describe the phase evolution of a signal appearing in just one detector.
Matched filter templates contain all the available information about the signal, whereas seedless clustering templates are a lossy, binned representation of the signal.
By throwing away information, the seedless clustering search is less sensitive than a matched filter search, but by throwing away information, it can simultaneously become more robust against waveform uncertainties and new physics.

A very general search with minimal assumptions may employ, e.g., a template bank of randomly generated B\'ezier curves~\cite{bezier}, which have been shown to do a reasonably good job of mimicking long-lived narrowband gravitational-wave signals~\cite{stochsky}.
However, one may equally well carry out a more specialized search, targeting a specific class of signals.
Given our present interest in CBC signals, we opt to work with a more specialized template bank consisting of parametrized chirps:
\begin{equation}\label{eq:f_of_t}
  f(t) = \frac{1}{2 \pi} \frac{c^{3}}{4 G M_\text{total}} \sum_{k=0}^7 p_k \tau^{-(3+k)/8} ,
\end{equation}
where
\begin{equation}
  \tau = \frac{\eta c^{3} (t_c -t)}{5 G M} .
\end{equation}
Here, $G$ is the gravitational constant and $M_\text{total}$ is the total mass of the binary.
The expansion coefficients $p_k$ can be found in \cite{PN}. 
Each chirp template is parametrized by two numbers: the coalescence time and the chirp mass.
(This is in contrast to B\'ezier curves, which are parametrized by six numbers.)
While technically, the waveform depends on the individual component masses, the main features of the signal can be well-approximated by only the chirp mass. Therefore, to reduce the parameter space by one variable, we employ the approximation that the individual component masses are equal.

The space of arbitrary long-lived gravitational-wave signals is very large, and so general algorithms, employing B\'ezier templates, use randomly generated numbers to span as much of the signal space as possible.
The space of CBC chirp signals is much smaller.
A search for binary neutron star signals with component masses of $1.4$--$3 M_\odot$ plateaus in sensitivity using just $50$ chirp mass bins.
The same search requires $825$ time bins for $\unit[660]{s}$ of data---a typical on-source window (in which the signal is assumed to exist) for a GW search triggered by a gamma-ray burst~\cite{0004-637X-760-1-12}.
Thus, the template density for a targeted CBC search is $\approx\unit[6.3\times10^4]{ks^{-1}}$ whereas a B\'ezier bank might require $\approx\unit[8\times10^8]{ks^{-1}}$~\cite{stochsky}; ($\unit[1]{ks}=\unit[1000]{s}$).
Since there are so many fewer templates in a chirp-template search (about four orders of magnitude fewer), it is computationally feasible to employ every template.

Of course, the previous calculation was for a targeted search in which the source location is previously determined, e.g., by an electromagnetic trigger or by a different search algorithm.
In \cite{stochsky}, we showed how the introduction of a phase factor can be used to allow to carry out an efficient all-sky search with seedless clustering.
This formalism is straightforwardly applied to our chirping templates.
For the CBCs detected by the LIGO Hanford-Livingston detector pair, it is sufficient to consider $40$ time delays, each corresponding to a ring on the sky.
Thus, even an all-sky search using seedless clustering to detect CBC signals can employ a relatively modest template density: $\approx\unit[2.5\times10^6]{ks^{-1}}$.

To estimate the computational cost of an all-sky seedless clustering search (with chirp-like templates), we carried out a benchmark study using a Kepler GK104s GPU and an 8-core Intel Xeon E5-4650 CPU.
Each job was allotted $\unit[8]{g}$ of memory.
The GPU was able to analyze $\unit[660]{s}$ of data in $\unit[48]{s}$, corresponding to a duty cycle of $\approx7\%$.
Using all eight cores, the CPU duty cycle was comparable; the job-by-job variability in run time is greater than the difference between GPUs and 8-core CPUs on average.

If we require background estimation at the level of $\text{FAP}=1\%$, it follows that a continuously running seedless clustering search with chirp-like templates can be carried out with just 8~GPUs (or 8-core CPUs).
(Background estimation at the level of $\text{FAP}=0.1\%$ would require 74~GPUs / 8-core CPUs.)
In reality, the duty cycle from coincident GW detectors may be $\approx50\%$, in which case these computing requirements are conservative by a factor of two.
Repeating the test for a targeted search (for which the source location is known), we obtained an 8-core CPU duty cycle of $2\%$, a factor of $3$ speed-up.
The targeted search run on GPUs does not run appreciably faster than the all-sky version.

\begin{figure*}[t]
 \includegraphics[width=3.5in]{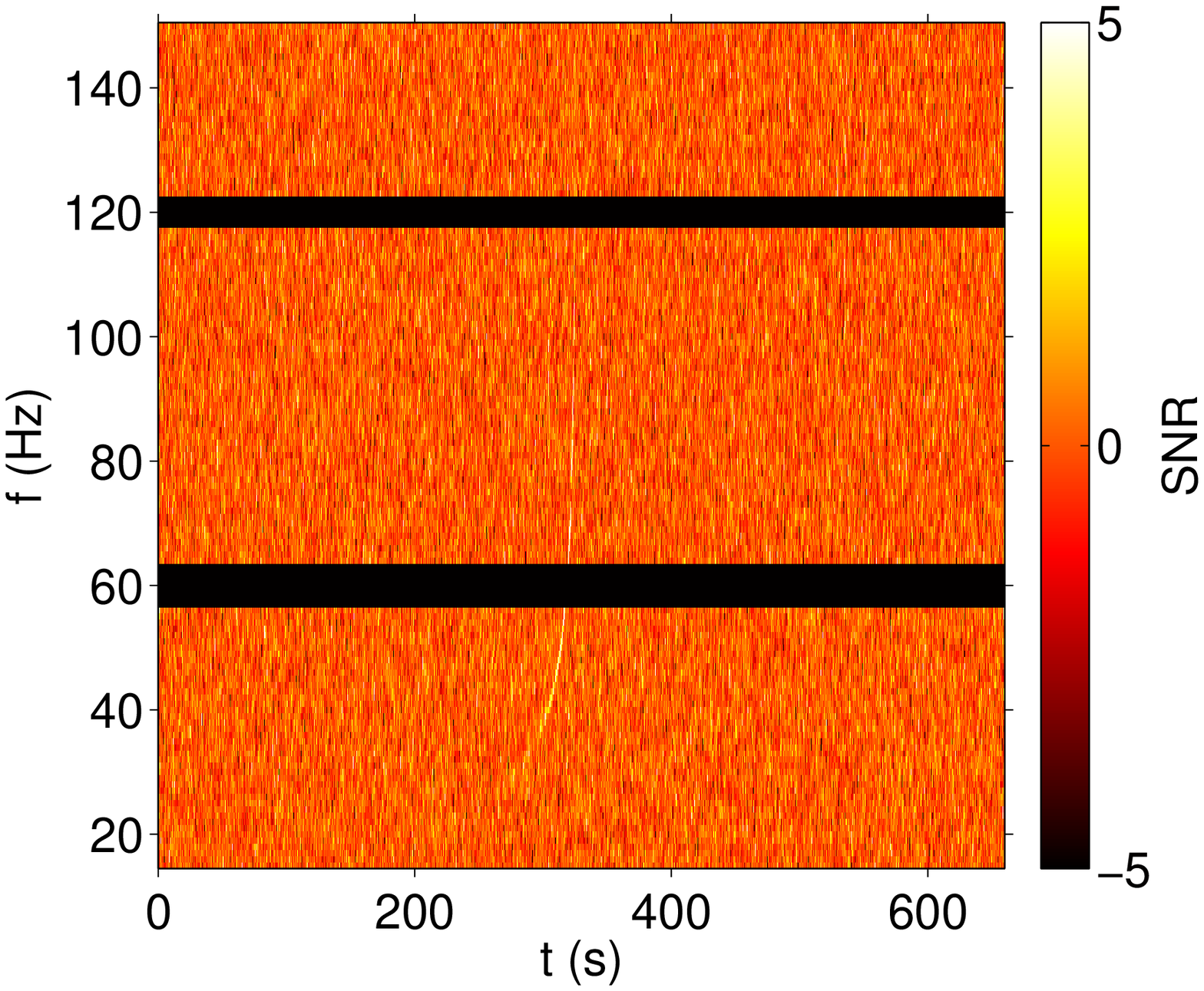}
 \includegraphics[width=3.5in]{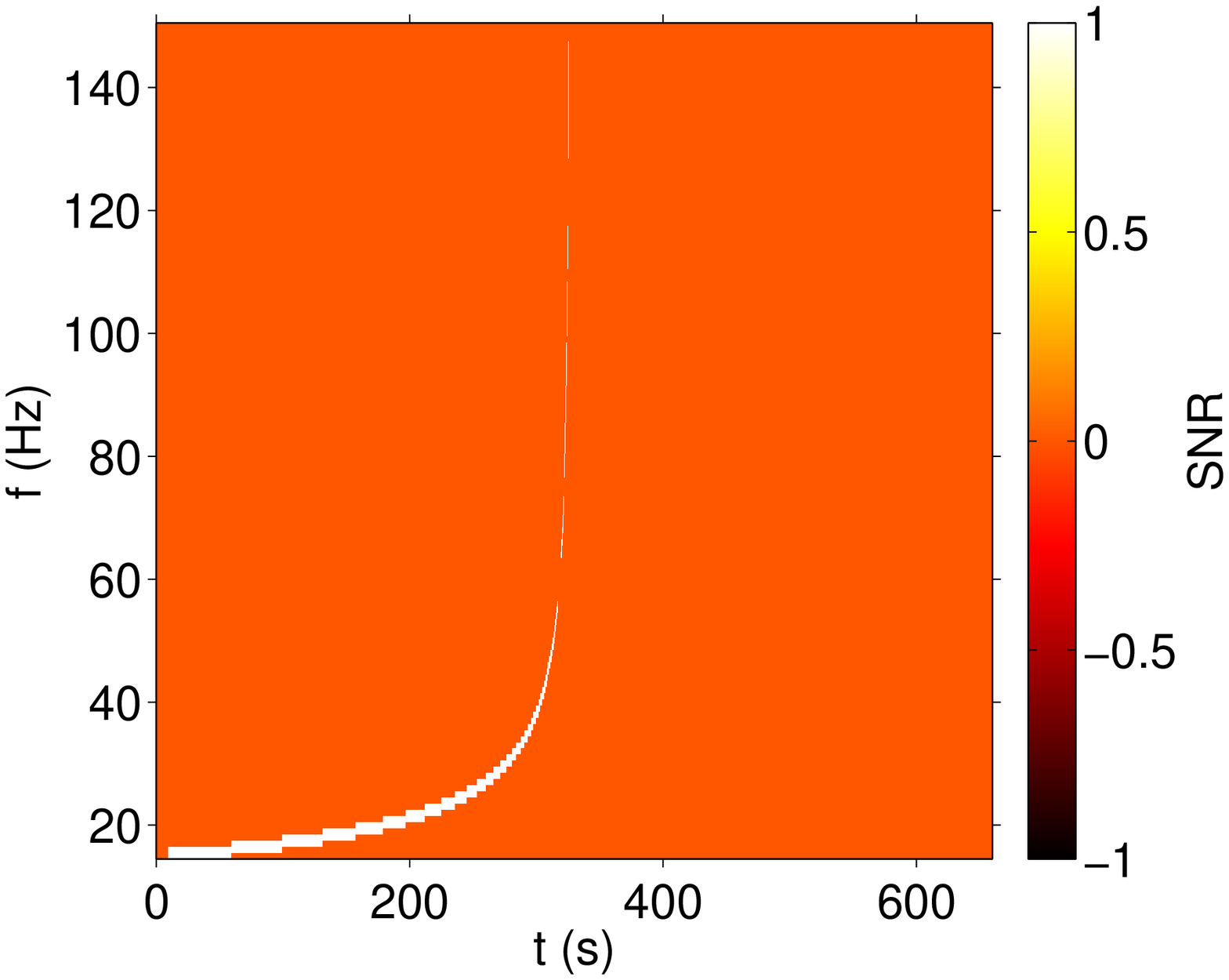}\\
 \caption{
   The plot on the left shows $\rho(t;f)$ for a simulated eccentric ($\epsilon = 0.2$) BNS signal injected on top of Monte Carlo detector noise.
   The component masses are $1.4 M_\odot$.
   The chirping signal appears as a faintly-visible track of lighter-than-average pixels.
   The horizontal lines are frequency notches to remove instrumental artifacts.
   On the right is the recovery obtained with seedless clustering.
   The signal is recovered with a $\text{FAP}<0.1\%$.
 }
 \label{fig:EBNS}
\end{figure*}

In addition to improved computational efficiency, there is another important advantage to be gained through the use of CBC templates compared to B\'ezier templates.
In \cite{stochsky}, we showed that quadratic B\'ezier curves do a mediocre job approximating CBC signals.
By adopting the chirping templates described in Eq.~\ref{eq:f_of_t}, we expect to capture more signal-to-noise ratio, and thereby extend the sensitive distance of the search.
An example of a weak BNS signal recovered with a seedless chirping template is shown in Fig.~\ref{fig:EBNS}.


\section{Sensitivity study}
In order to determine the sensitivity of seedless clustering with chirp templates, we perform a sensitivity study with Monte Carlo noise.
We assume Gaussian noise consistent with the design sensitivity of Advanced LIGO.
Following~\cite{0004-637X-760-1-12}, we assume that an external trigger has predicted that the signal exists in a $\unit[660]{s}$ on-source window.
For each trial, we search for a chirp signal four different ways: using B\'ezier curves and a known sky location (BK), using B\'ezier with unknown sky location (BU), using chirp templates with a known sky location (CK), and using chirp templates with an unknown sky location (CU).

The first step of the sensitivity study is background estimation.
We perform many trials to estimate the distribution of $\text{SNR}_\text{tot}$ for noise.
We generate separate noise distributions for all four search variations (BK, BU, CK, CU).
Using these noise distributions, we determine the value of $\text{SNR}_\text{tot}$ (for each search variation), which corresponds to a false alarm probability (FAP) of $0.1\%$.

The next step is to determine the distance to which different signals can be detected with $\text{SNR}_\text{tot}$ sufficient for a detection with $\text{FAP}<0.1\%$.
We add GW signals to realizations of detector noise.
Each injected signal is injected with an optimal sky location and an optimal source orientation.
We define the ``sensitive distance'' as the distance at which $50\%$ of the singals are recovered with $\text{FAP}<0.1\%$.
We consider $14$ CBC waveforms with component masses ranging from $1.4$--$3 M_\odot$.
Of these waveforms, eight characterize eccentric systems and three characterize systems where one or more object has a large dimensionless spin:
\begin{equation}\label{eq:spin}
  a \equiv c J / G m^2 .
\end{equation}
Here $J$ is the angular momentum and $m$ is the component mass.

Non-eccentric waveforms are generated using a SpinTaylorT4 approximation.
Eccentric waveforms are generated using {\tt CBWaves}, which employs all the contributions that have been worked out for generic eccentric orbits up to 2PN order \cite{CBWaves}.
The parameters for each waveform are give in Table~\ref{tab:Results}.

\begin{table*}[t]
\begin{tabular}{|c|c|c|c|c|c|c|c|c|c|c|c|c|}
\hline
Waveform & $m_1$ & $m_2$ & $a_1$ & $a_2$ & $\epsilon$ & $t_\text{dur}$ (s) & $D_\text{BK}$ & $D_\text{BU}$ & $D_\text{CK}$ & $D_\text{CU}$ \\
\hline\hline
BNS 1 & 1.4 & 1.4 & 0 & 0 & 0 & 170 & 160 & 130 & 190 & 190 \\\hline
NSBH 1 & 3.0 & 1.4 & 0 & 0 & 0 & 96 & 200 & 200 & 290 & 290\\\hline
NSBH 2 & 3.0 & 1.4 & 0.95 & 0 & 0 & 97 & 220 & 180 & 320 & 320\\\hline
BBH 1 & 3.0 & 3.0 & 0 & 0 & 0 & 54 & 330 & 330 & 470 & 470\\\hline
BBH 2 & 3.0 & 3.0 & 0.95 & 0 & 0 & 55 & 320 & 270 & 470 & 470\\\hline
BBH 3 & 3.0 & 3.0 & 0.95 & 0.95 & 0 & 55 & 320 & 260 & 470 & 470\\\hline
BBH 4 & 5.0 & 5.0 & 0 & 0 & 0 & 42 & 620 & 560 & 750 & 750\\\hline
BBH 5 & 5.0 & 5.0 & 0.95 & 0 & 0 & 42 & 680 & 680 & 750 & 750\\\hline
BBH 6 & 5.0 & 5.0 & 0.95 & 0.95 & 0 & 43 & 750 & 680 & 830 & 830\\\hline
EBNS 1 & 1.4 & 1.4 & 0 & 0 & 0.2 & 120 & 150 & 120 & 180 & 180\\\hline
EBNS 2 & 1.4 & 1.4 & 0 & 0 & 0.4 & 224 & 150 & 120 & 160 & 160\\\hline
ENSBH 1 & 3.0 & 1.4 & 0 & 0 & 0.2 & 69 & 180 & 180 & 290 & 290\\\hline
ENSBH 2 & 3.0 & 1.4 & 0 & 0 & 0.4 & 127 & 180 & 160 & 240 & 240\\\hline
ENSBH 3 & 3.0 & 1.4 & 0 & 0 & 0.6 & 237 & 180 & 160 & 240 & 240\\\hline
EBBH 1 & 3.0 & 3.0 & 0 & 0 & 0.2 & 40 & 270 & 220 & 320 & 320\\\hline
EBBH 2 & 3.0 & 3.0 & 0 & 0 & 0.4 & 70 & 220 & 200 & 240 & 240\\\hline
EBBH 3 & 3.0 & 3.0 & 0 & 0 & 0.6 & 128 & 220 & 180 & 220 & 220\\\hline
\end{tabular}
\caption{
  Sensitive distances for different waveforms (assuming optimal sky location and source orientation) given the design sensitivity of Advanced LIGO~\cite{obs_scen}.
  Each row represents a different waveform: BNS=``binary neutron star,'' NSBH=``neutron star black hole binary,'' BBH=``binary black hole.
  A waveform beginning with an ``E'' is eccentric.
  The columns marked $m_1$  and $m_2$ give the component masses in units of $M_\odot$.
  The columns marked $a_1$ and $a_2$ give the component spins; see Eq.~\ref{eq:spin}.
  The next columns list the ellipticity $\epsilon$ and the waveform duration in seconds.
  The final four columns list the ($\text{FAP}=0.1\%$, $\text{FDP}=50\%$) detection distance (in Mpc) for B\'ezier templates with known sky location (BK),  B\'ezier templates with unknown sky location (BU), chirp-like templates with known sky location (CK), and chirp-like templates with unknown sky location (CU).
}
\label{tab:Results}
\end{table*}

\section{Results}\label{sec:Results}
The results of our sensitivity study are summarized in Table~\ref{tab:Results}.
There are a number of interesting trends.
First, while all of the (known direction) seedless clustering distances are astrophysically interesting, the chirping templates perform consistently better than the B\'ezier templates.
The ratio of detection distance for chirping templates / B\'ezier templates ranges from $100$--$161\%$ with a mean of $131\%$.
The average ratio of sensitive volumes is $240\%$.
This is comparable to the gain in sensitivity for a matched filter search to be had through the inclusion of spin~\cite{SteveP:Thesis:2014}, indicating that while it is certainly advantageous to use chirping templates, the B\'ezier templates do surprisingly well.

We find no significant difference in the chirping-template detection distance between systems that do or do not contain spin.
The similarity in the sensitivity distances between the non-spinning and spinning cases indicates that the spins do not effect the signal morphology in a significant enough way to deviate from the non-spinning track. 
The advantage of chirping templates appears to increase slightly for spinning systems.
Finally, we observe no loss in sensitivity going from the targeted CK search to the all-sky CU search.
Evidently, the increase in signal space from the additional parameter of sky location is not sufficient to meaningfully affect the background distribution.

Highly eccentric signals generally have a longer duration than than those with low or no eccentricity. The sensitivity (using both B\'ezier templates and chirp-like templates) decreases with eccentricity. There is a similar advantage in distance of the chirp-like templates over the B\'ezier templates at low eccentricity. This benefit decreases slightly as eccentricity increases.
This is due to the breakdown of the circular binary approximation.
The breakdown becomes more pronounced at higher eccentricities, as one would expect.

Using matched filtering, Advanced LIGO, operating at design sensitivity, is expected to reliably detect BNS (with optimal orientation and sky location) out to distances of $\unit[450]{Mpc}$~\cite{obs_scen}, $2.4\times$ further than the seedless clustering detection distance quoted here.
It follows that $\approx8\%$ of the events detected by matched filtering will produce a $\text{FAP}<0.1\%$ signature when followed up with seedless clustering.
Given a realistic astrophysical rate of $\unit[40]{yr^{-1}}$ BNS detections by Advanced LIGO~\cite{rates}, this implies that we can expect to confirm $\approx 3$ events per year of science data using seedless clustering.
The use of expanded template banks that include waveforms with spin will allow matched filtering searches to observe to comparable distances as the non-spinning case \cite{PhysRevD.86.084017}; therefore the follow-up numbers will be similar to the above.
In the event that circular template banks are used to search for eccentric signals, there will be a non-negligable loss in sensitivity for these searches. Huerta and Brown estimate signal-to-noise ratio loss factors of about 0.5 and 0.2 for BNS systems with eccentricities of 0.2 and 0.4 respectively \cite{PhysRevD.87.127501}. This would bring the matched filtering sensitivity distances of these signals to $\unit[225]{Mpc}$ and $\unit[90]{Mpc}$; therefore seedless clustering may provide further opportunities for observing these types of signals.

In addition to providing confirmation of these loudest CBC events, seedless clustering will provide a safety net by potentially detecting events missed due to waveform error, data-processing subtleties, and/or new physics.

\section{Conclusions}\label{sec:Conclusion}
Seedless clustering provides a computationally efficient tool for the follow-up and detection of compact binary coalescences.
While seedless clustering is expected to be less sensitive than matched filtering, it provides a number of useful features including independent verification, visualization of the gravitational-wave signal, the ability to catch corner-case signals, and robustness to both waveform uncertainty and existence of new physics.

We compared a specially tuned implementation of seedless clustering, optimized for compact binary coalescences, to a more generic search using B\'ezier curves.
We find that the CBC-tuned search can expand the sensitive volume by as much as a factor of $4.2\times$ depending on the waveform (a factor of two on average) compared to the generic B\'ezier search.
Perhaps more importantly, the tuned search requires $10^4$ fewer templates per unit of time, allowing for a significantly faster search.


There are a number of potential improvements to the algorithm worth exploring. 
It may be possible to improve the implementation of seedless clustering described here by more optimally weighting different time-frequency bins based on the known waveform. 
It is also worth exploring the effect of only using equal mass templates to recover potentially non-equal mass signals. It is possible that for cases with larger mass ratios than those considered here, it may be necessary to relax this assumption in order to reconstruct a majority of the signal.
Another possibility for improvement is the implementation of a better parametrization for the eccentric waveforms.
Unlike for circular binaries, there is, at present, no closed expression for the phase evolution of a binary with arbitrary eccentricity.
A more effective parametrization is likely to capture more signal-to-noise ratio, thereby extending the sensitive range.
Finally, it will be useful to carry out a systematic comparison of seedless clustering with matched filtering pipelines and using non-Gaussian noise.


\section{Acknowledgments}
MC is supported by National Science Foundation Graduate Research Fellowship Program, under NSF grant number DGE 1144152.
ET is a member of the LIGO Laboratory, supported by funding from United States National Science Foundation.
NC's work was supported by NSF grant PHY-1204371.
LIGO was constructed by the California Institute of Technology and Massachusetts Institute of Technology with funding from the National Science Foundation and operates under cooperative agreement PHY-0757058.
This paper has been assigned LIGO document number P1400110-v5.

\appendix

\begin{figure*}[t]
 \includegraphics[width=6.0in,height=2.9in]{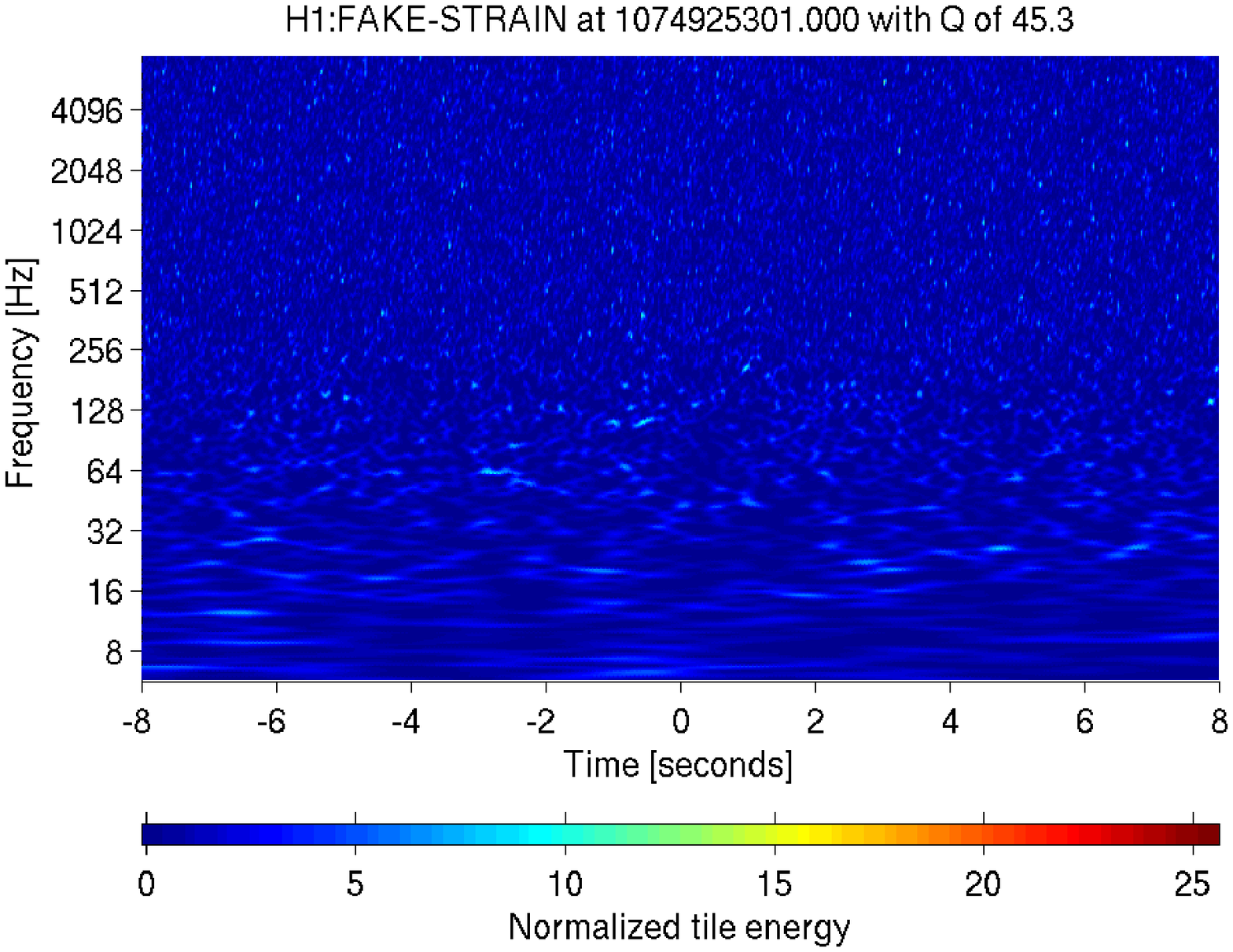}
 \includegraphics[width=6.0in,height=2.9in]{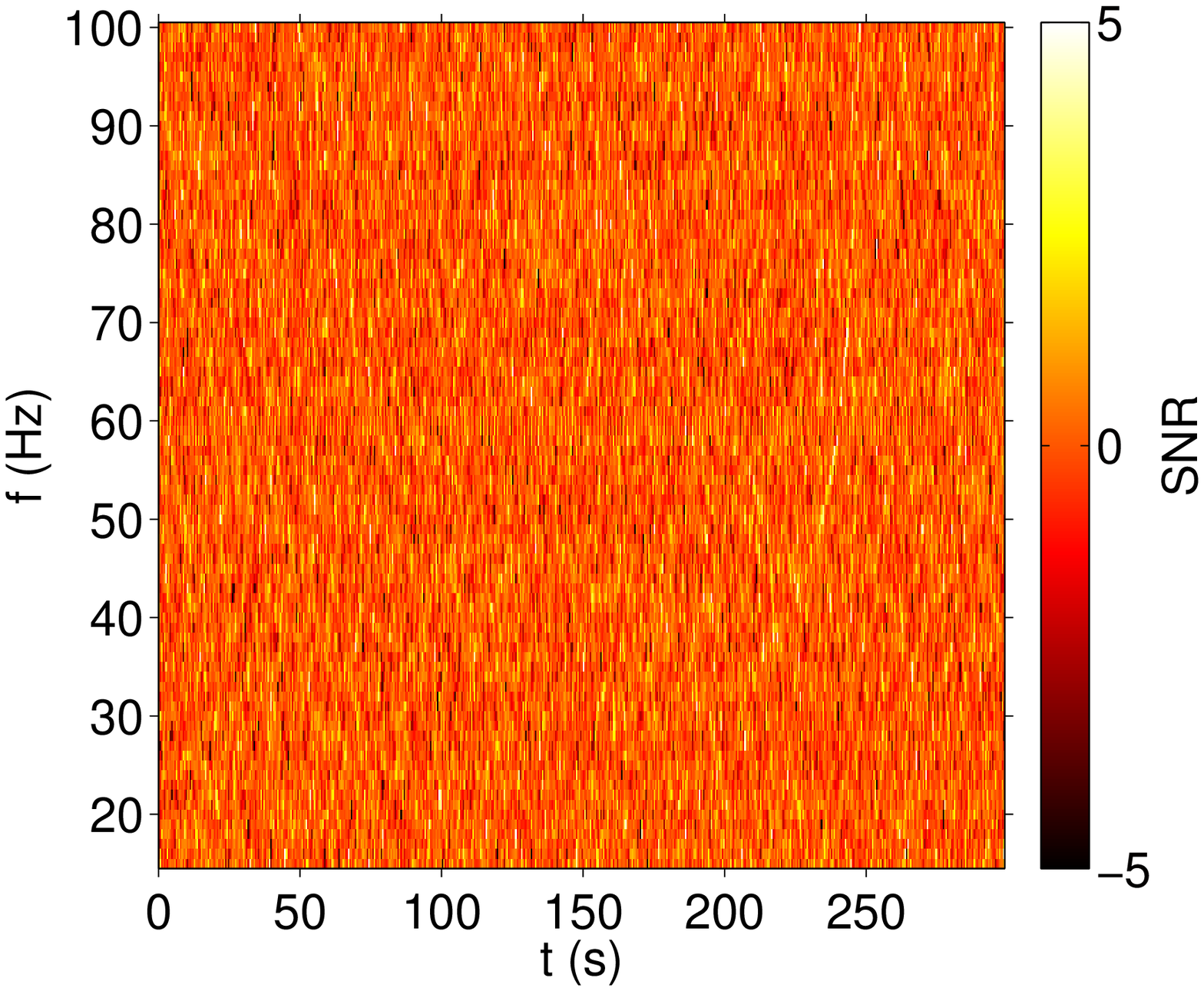}\\
 \hspace{0.2in}
 \includegraphics[width=6.2in,height=2.9in]{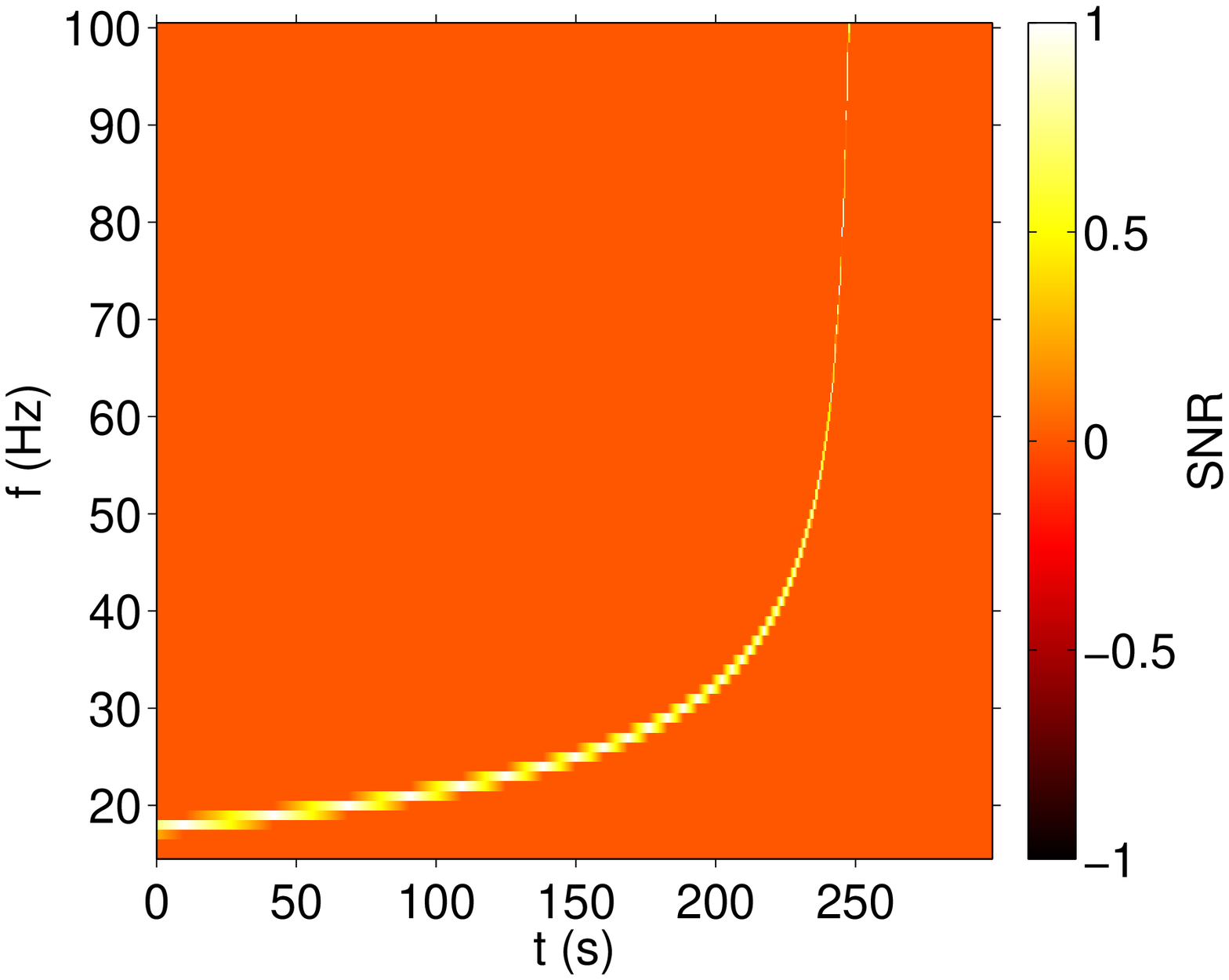}\\
 \caption{
   Data from a recent LIGO engineering run, in which data from a LIGO subsystem---not sensitive to GW strain---is recolored to produce semi-realistic detector noise.
   A simulated binary neutron star signal has been added to the data.
   The top plot shows a wavelet transform of single-detector auto-power~\cite{JamieR:Thesis:2011}, currently in wide use for diagnostics.
   The injected waveform, which ends at $t=0$, is difficult to make out by eye.
   The middle plot shows a spectrogram of $\rho(t;f)$.
   The injection, though faint, is visible between 200-$\unit[250]{s}$.
   The bottom plot shows the reconstructed track using seedless clustering; $\text{FAP}<0.1\%$.
 }
 \label{fig:ER5Omega}
\end{figure*}



\bibliography{stochtrack_cbc}

\end{document}